\documentclass[aps,prb,reprint,a4paper]{revtex4}

\usepackage{amsmath,amssymb,mathrsfs}
\usepackage[dvips]{graphicx}

\newcommand{\bs}[1]{\boldsymbol{#1}}
\newcommand{\ket}[1]{\left|#1\right\rangle}
\newcommand{\bra}[1]{\left\langle#1\right|}

\newcommand{\ii}{\text{i}}
\newcommand{\be}{\begin{equation}}
\newcommand{\ee}{\end{equation}}
\newcommand{\bea}{\begin{eqnarray}}
\newcommand{\eea}{\end{eqnarray}}

\allowdisplaybreaks[4]

\begin{document}
\title{Dynamics in the Ising field theory after a quantum quench}
\author{Dirk Schuricht}
\affiliation{Institute for Theory of Statistical Physics, 
RWTH Aachen, 52056 Aachen, Germany}
\affiliation{JARA-Fundamentals of Future Information Technology}
\author{Fabian H.L. Essler}
\affiliation{The Rudolf Peierls Centre for Theoretical Physics, 
University of Oxford,\\ 1 Keble Road, OX1 3NP, Oxford, United Kingdom}
\date{\today}
\pagestyle{plain}

\begin{abstract}
We study the real-time dynamics of the order parameter
$\langle\sigma(t)\rangle$ in the Ising field theory after a
quench in the fermion mass, which corresponds to a quench in the
transverse field of the corresponding transverse field Ising
chain. We focus on quenches within the ordered phase. The long-time
behaviour is obtained analytically by a resummation of the leading
divergent terms in a form-factor expansion for
$\langle\sigma(t)\rangle$. Our main result is the development
of a method for treating divergences associated with working directly
in the field theory limit. We recover the scaling limit of the
corresponding result by Calabrese et al. [Phys. Rev. Lett. {\bf
  106}, 227203  (2011)], which was obtained for the lattice model. Our
formalism generalizes to integrable quantum quenches in other
integrable models. 
\end{abstract}

\maketitle

\section{Introduction}
Recent experimental advances have made it possible to study the
non-equilibrium dynamics of trapped cold atomic gases
\cite{Greiner02,kww-06}. A key feature of these
systems is that they are only weakly coupled to their
environments, which makes it possible to study non-equilibrium
dynamics in essentially isolated systems. This has led to an intense
theoretical effort to address  fundamental questions\cite{rev}
regarding the non-equilibrium dynamics of many-body quantum systems.

One issue of particular interest concerns the role played by
dimensionality and conservation laws. As shown by the ``quantum
Newton's cradle'' experiments of Kinoshita et al. \cite{kww-06},
quasi-one dimensional condensates exhibit behaviour that is
dramatically different from two and three dimensional ones. In
particular, it was observed that the late-time behaviour cannot be
described in terms of an effective temperature: the systems does
not ``thermalize'' \cite{thermalization}. One possible explanation\cite{kww-06} 
for this behaviour is that the experimental system is close to being
integrable. This has engendered a vigorous research effort aimed at
clarifying the role played by quantum integrability in the late time
and stationary state behaviour in non-equilibrium dynamics
\cite{gg,cc-06,gge_various,fm-10,Gritsev,CEF-11}.  

A simple and attractive way of inducing non-equilibrium evolution is by
means of a {\it quantum quench}. One prepares a system in the ground
state of a given Hamiltonian $H(h_0)$, where $h_0$ is an experimentally
tuneable parameter such as a magnetic field or an interaction
strength. At time $t=0$ the parameter $h_0$ is then changed
instantaneously from $h_0$ to $h$, and at subsequent times the system
evolves according to the quantum dynamics induced by the new
Hamiltonian $H(h)$. One of the main models studied in the context of
quantum quenches has been the transverse field Ising chain\cite{mc,rsms-08,CEF-11}. 
This is on the one
hand because the model has a free fermion representation which makes
analytical progress possible. On the other hand the model is the
simplest paradigm of a quantum phase transition and therefore is an
ideal testing ground for questions relating to non-equilibrium
evolution in the vicinity of quantum critical points. The stationary
and late-time behaviour of correlation functions in the transverse field Ising chain after a
quantum quench has recently been determined analytically by Calabrese,
Essler and Fagotti\cite{CEF-11}. 

Analyzing quantum quenches in \emph{interacting} integrable models is
difficult\cite{Spyros} and remains a largely open challenge, although
important progress has been made by combining numerical and
integrable model techniques \cite{Gritsev}. A special role is
played by \emph{integrable} quenches in integrable quantum field
theories. These are characterized as follows. As shown by Calabrese
and Cardy \cite{cc-06} the quench problem can be mapped to an
equivalent theory defined in a strip geometry. The initial state plays
the role of a boundary condition, and for an integrable quench this
boundary condition does not spoil the integrability of the
theory. Hence, for these special initial states one can use methods of
integrable quantum field theory\cite{Smirnov92book,Lukyanov95} with boundaries
\cite{GhoshalZamolodchikov94} to analyze the time evolution of observables. 
An important step in this direction was taken by Fioretto and
Mussardo\cite{fm-10}, who considered the stationary state behaviour of
one-point functions in integrable quenches and in particular in the
Ising field theory. A serious complication that arises in the field
theory limit is that singularities associated with kinematical poles
appear. This problem is particularly acute for two-point correlators
and is closely related to the one encountered when calculating 
finite-temperature dynamical correlation functions in integrable models
\cite{ek09,KormosPozsgay10,finiteT}. To date two general ways of dealing
with these singularities have been developed. The first
\cite{ek09,KormosPozsgay10,finiteT} is to use a finite-volume regularization for
matrix elements\cite{takacs}, while the second is a subtraction scheme
that works directly in the infinite volume\cite{ek09}. The aim of the
present work is to apply these methods to the problem of quench
dynamics in the ordered phase of the Ising field theory. Regulating
the theory in a finite volume reduces all calculations to a particular
limit of the analysis for the lattice Ising model\cite{CEF-11,CEF-12}
and we therefore do not report any details here. We focus on the
infinite-volume regularization proposed in Ref.~[\onlinecite{ek09}] and
apply it to the quench problem at late, finite times. This requires a
significant generalization of the regularization procedure, which
constitutes the main result reported here. In forthcoming work we will
apply this method to a quench in the sine-Gordon model.

The outline of this paper is as follows: In Section~\ref{sec:QIC} we introduce the 
Ising field theory as the scaling limit of the transverse field Ising chain. 
In Sec.~\ref{sec:quench} we discuss quenches in the fermion mass of the 
field theory, which corresponds to a quench in the transverse field of the related Ising 
chain. In Sec.~\ref{sec:method} we develop a new method to calculate the time 
evolution of correlation functions in integrable field theories, which constitutes the
main result of our work. In Sec.~\ref{sec:results} we apply this method to the 
one-point function of the order parameter field, which relaxes exponentially to zero
as shown in Eqs.~\eqref{eq:sigmat} and~\eqref{eq:relaxationrate}. In 
Sec.~\ref{sec:FM} we discuss the relation of quenches in the fermion mass to the 
extrapolation time regularization introduced in Ref.~[\onlinecite{fm-10}] and conclude in
Sec.~\ref{sec:conclusion}. Technical details of the derivations have been moved to 
the appendices.

\section{Quantum Ising chain}\label{sec:QIC}
We start with the transverse field Ising chain
\begin{equation}
\label{eq:Isinglattice}
H_\mathrm{latt}=-J\sum_i \bigl(\sigma^\mathrm{z}_i\,\sigma^\mathrm{z}_{i+1}
+h \,\sigma^\mathrm{x}_i\bigr).
\end{equation}
Here $\sigma^\mathrm{x}$ and $\sigma^\mathrm{z}$ are the Pauli matrices and
$J$ is the exchange energy. The Hamiltonian \eqref{eq:Isinglattice} is invariant 
under the $\mathbb{Z}_2$-transformation 
$\sigma_i^\mathrm{x}\rightarrow\sigma_i^\mathrm{x}$,
$\sigma_i^\mathrm{z}\rightarrow-\sigma_i^\mathrm{z}$.  For $h<1$ this
symmetry is broken, the order parameter field $\sigma_i^\mathrm{z}$ takes a
non-zero expectation value, and the ground state is two-fold degenerate. 
On the other hand, for $h>1$ the system possesses a unique ground state
and the expectation value of the order parameter field vanishes. The two phases
are separated by a quantum critical point at $h=1$.
At small deviations from criticality, $|h-1|\ll 1$, one can pass to the
scaling limit~\cite{ItzyksonDrouffe89vol1} ($a_0$ is the lattice spacing) 
\be
J\to\infty,\quad h\to 1,\quad a_0\to 0,
\label{scalingI}
\ee
while keeping fixed both the gap $M$ and the velocity $v$
\be
2J|1-h|=M,\quad 2Ja_0=v.
\label{scalingII}
\ee
The order parameter in the scaling limit must be defined as
\be
\sigma(x)\propto(1-h^2)^{-\frac{1}{8}}\sigma^\mathrm{z}_i,
\ee
where $x=na_0$. It is customary to choose the normalization of the
field $\sigma(x)$ such that 
\be
\lim_{x\to 0}\ \langle
0|\sigma(x)\sigma(0)|0\rangle=\frac{1}{|x|^\frac{1}{4}},
\ee
which implies
\be
\sigma^\mathrm{z}_i\rightarrow 2^{1/24}e^{1/8}{\cal
  A}^{-3/2}a_0^{1/8}\sigma(x),
\ee
with Glaisher's constant
\be
{\cal A}=1.28242712910062...
\ee
The Hamiltonian in the scaling limit reads
\begin{equation}
H =\int_{-\infty}^\infty \frac{dx}{2\pi}\left[ \frac{\ii v}{2}
(\psi\partial_x\psi-{\bar\psi}\partial_x\bar\psi) -
\ii M{\bar\psi}\psi\right],
\label{Hscaling}
\end{equation} 
where $\psi$ and $\bar{\psi}$ are the two components of a Majorana fermion.
The model \eqref{Hscaling} 
is conformally invariant at the critical point $M=0$ (see for example
Ref.~[\onlinecite{DiFrancescoMathieuSenechal97}]). In the ordered phase, which
we will consider throughout this paper, the mass is positive. 

An important notion is the mutual semi-locality of
operators\cite{YurovZamolodchikov91,Smirnov92book,Lukyanov95,Delfino04}. 
This is most easily established by defining complex coordinates
$z=\tau+\ii x$, $\bar{z}=\tau-\ii x$ and then considering the operator product 
$O_1(\tau,x)\,O_2(0,0)=O_1(z,\bar{z})\,O_2(0,0)$. If we take $O_1$
counterclockwise around $O_2$ in the plane, i.e. we perform the
analytic continuation $z\rightarrow e^{2\pi\ii}z$, $\bar{z}\rightarrow
e^{-2\pi\ii}\bar{z}$, the operators $O_1$ and $O_2$ are said to be
\emph{mutually semi-local} if 
\begin{equation}
\label{eq:defsemilocality}
O_1(e^{2\pi\ii}z,e^{-2\pi\ii}\bar{z})\,O_2(0,0)=
l_{O_1O_2}\,O_1(z,\bar{z})\,O_2(0,0).
\end{equation}
The phase $l_{O_1O_2}$ is called the semi-locality factor. The two fields are
mutually local if $l_{O_1O_2}=1$. Semi-locality is the mildest form of
non-locality, in general the right-hand side of \eqref{eq:defsemilocality} may
be more complicated. The mutual semi-locality factor of the spin and
disorder operators can be extracted from their operator product
expansion~\cite{DiFrancescoMathieuSenechal97,Delfino04} 
\begin{equation}
\sigma(z,\bar{z})\,\mu(0,0)\sim
\frac{1}{\sqrt{2}\,|z|^{1/4}}\,\bigl[e^{\ii\pi/4}\sqrt{z}\,\psi(0)
+e^{-\ii\pi/4}\sqrt{\bar{z}}\,\bar{\psi}(0)\bigr].
\label{eq:OPE}
\end{equation}
This implies that when taking $\sigma$ once around $\mu$
one obtains an extra minus sign, i.e. $l_{\sigma\mu}=-1$.
In the same way one finds
$l_{\psi\mu}=l_{\bar{\psi}\mu}=
l_{\psi\sigma}=l_{\bar{\psi}\sigma}=-1$. On the
other hand, the disorder field $\mu$ is local with respect
to itself.  

We use the disorder field $\mu$ as fundamental field creating the 
excitations. This implies that the fundamental excitations are viewed as bosons.
We denote the corresponding annihilation and creation operators by $A(\theta)$ 
and $A^\dagger(\theta)$ respectively. They fulfil the Faddeev-Zamolodchikov 
algebra\cite{ZamolodchikovZamolodchikov79}
\begin{eqnarray}
A(\theta_1)A(\theta_2)&=&SA(\theta_2)A(\theta_1),\nonumber\\*
A^\dagger(\theta_1)A^\dagger(\theta_2)&=&
SA^\dagger(\theta_2)A^\dagger(\theta_1),\label{eq:Aalgebra}\\*
A(\theta_1)A^\dagger(\theta_2)&=&2\pi\delta(\theta_1-\theta_2)
+SA^\dagger(\theta_2)A(\theta_1),\nonumber
\end{eqnarray}
with the scattering matrix $S=-1$. The basis of scattering states can now be constructed by
\begin{equation}
  \label{eq:defA}
  \ket{\theta_1,\ldots,\theta_n}=
  A^\dagger(\theta_1)\ldots A^\dagger(\theta_n)\ket{0},
\end{equation}
where the vacuum state $\ket{0}$ is defined by $A(\theta)\ket{0}=0$. 
The energy and momentum of the scattering states are expressed in
terms of the rapidities $\theta_i$ as
\begin{equation}
E=M\sum_{i=1}^n\cosh\theta_i,\quad P=\frac{M}{v}\sum_{i=1}^n\sinh\theta_i.
\end{equation}

In this article we study the one-point function of the order parameter field $\sigma$. 
The relevant matrix elements (form factors) in the ordered phase are given 
by~\cite{IsingFF,YurovZamolodchikov91,Delfino04} 
\begin{equation}
\label{eq:sigmaff}
f(\theta_1,\ldots,\theta_{2n})=
\bra{0}\sigma\ket{\theta_{1},\ldots,\theta_{2n}}=
\ii^n\,\bar{\sigma}\prod_{\substack{i,j=1\\ i<j}}^{2n}
\tanh\frac{\theta_i-\theta_j}{2},
\end{equation}
where 
\be
\bar{\sigma}=\bra{0}\sigma\ket{0}=2^{1/12}e^{-1/8}{\cal A}^{3/2}\left(\frac{M}{v}\right)^{1/8}.
\ee

\section{Quench in the fermion mass}\label{sec:quench}
We now consider a sudden change of the transverse field in
\eqref{eq:Isinglattice} at time $t=0$ from $h_0$
to $h$. This quench has been studied previously by several
authors \cite{mc,rsms-08,fm-10}. Most
importantly, in Refs.~[\onlinecite{CEF-11,CEF-12}] the time evolution of
both the one-point and two-point function of the order parameter after a
quench was determined analytically. One of the methods developed in
Refs.~[\onlinecite{CEF-11,CEF-12}] is a form-factor approach for the lattice
model. In the following we consider the time evolution of the one-point
function directly in the scaling limit (\ref{scalingI}),
(\ref{scalingII}). As we have mentioned before, our key objective is
to generalize the form-factor approach to quantum field theories in
order to analyze integrable quenches in interacting systems such as
the sine-Gordon model. However, a second interesting issue is related
to commutativity of limits: a priori it is unknown whether a quench in
the scaling limit is the same as the scaling limit of a quench. 
We will come back to this question in Secs.~\ref{sec:longtime}
and~\ref{sec:conclusion}.
In the following we resolve this question for the particular case of the
one-point function of the order parameter in the ferromagnetic phase of
the Ising model.

In the field theory \eqref{Hscaling} the quench in the transverse field
corresponds to a quench in the fermion mass, i.e. at time $t=0$ we
switch from $M_0$ to $M$. The time evolution for $t>0$ is governed by
\eqref{Hscaling}, while the initial state can be expressed in terms of
the eigenstates of \eqref{Hscaling} as \cite{rsms-08,Spyros} 
\begin{equation}
  \label{eq:initialstate}
  \ket{\Psi_0}=\exp\left(\int_0^\infty\frac{d\xi}{2\pi}
    K_\text{q}(\xi)A^\dagger(-\xi)A^\dagger(\xi)\right)\ket{0},
\end{equation}
where 
\begin{equation}
K_\text{q}(\xi)=\ii\tan\left[\frac{1}{2}\text{arctan}(\sinh\xi)
-\frac{1}{2}\text{arctan}\left(\frac{M}{M_0}\sinh\xi\right)\right]
\equiv \ii\,\hat{K}_\text{q}(\xi).
\label{eq:quenchK}
\end{equation}
We note that the quench matrix satisfies $K_\text{q}(\xi)=S\,K_\text{q}(-\xi)=-K_\text{q}(-\xi)$
and that $\hat{K}_\text{q}(\xi)\in\mathbb{R}$ for $\xi\in\mathbb{R}$. Furthermore, for any 
finite initial mass, $M_0<\infty$, the integral
\begin{equation}
  \int_0^\infty\frac{d\xi}{2\pi}\,\big|K_\text{q}(\xi)\big|^2
\end{equation}
is convergent. We note the similarity of the initial state \eqref{eq:initialstate} with the boundary 
state~\cite{GhoshalZamolodchikov94} introduced in the context of
integrable field theories with boundaries, which can be used to study
the physical properties of systems with defects or impurities\cite{SE07,impurities}. 
Starting from the initial state \eqref{eq:initialstate} we calculate the time evolution of the 
one-point function of the order parameter field,
\begin{equation}
\langle\sigma(t)\rangle\equiv
\frac{\bra{\Psi_0}\sigma(t)\ket{\Psi_0}}
{\bra{\Psi_0}\Psi_0\rangle},\quad
\sigma(t)=e^{\ii Ht}\sigma e^{-\ii Ht}.
\label{eq:At}
\end{equation}

\section{Method}\label{sec:method}
The strategy to calculate the one-point function  \eqref{eq:At} after
the quench is as follows: (i) We formally
expand~\cite{fm-10,CEF-11,KormosPozsgay10}  
the numerator and denominator in powers of the quench matrix 
$K_\text{q}$. (ii) We evaluate each term in these expansions using a
combined approach based on a regularization of the appearing form
factors following Smirnov~\cite{Smirnov92book} as well as the
$\kappa$-regularization recently introduced in the study of dynamical
correlation functions at finite temperatures~\cite{ek09}. (iii) In the
resulting expression the singularities in the numerator and
denominator, which are due to the infinite volume of the model
\eqref{Hscaling}, cancel each other. In particular we show by explicit 
calculation up to $\mathcal{O}(K_\text{q}^4)$ that this procedure
yields well defined results which agree with a finite-volume
regularization. (iv) Finally the resulting series has to be 
resummed~\cite{CEF-11} in order to obtain a well-defined long-time limit.
The calculation of two-point functions follows the same lines,
although the explicit expressions become considerably more complicated.

\subsection{Formal expansion}
The first step in the calculation of \eqref{eq:At} is the formal
expansion of both the numerator and the denominator in powers of the
quench matrix $K_\text{q}$. This expansion yields for 
the numerator
\begin{eqnarray}
  \bra{\Psi_0}\sigma(t)\ket{\Psi_0}&=&
  \sum_{m,n=0}^\infty\int_{0}^\infty
  \frac{d\xi_1'\ldots d\xi_m'}{m!(2\pi)^m}
  \frac{d\xi_1\ldots d\xi_n}{n!(2\pi)^n}
  \prod_{i=1}^m K_\text{q}(\xi_i')^*\,\prod_{j=1}^n K_\text{q}(\xi_j)\,
  e^{2M\ii t\sum_i\cosh\xi_i'}\,e^{-2M\ii t\sum_j\cosh\xi_j}\nonumber\\*
  & &\hspace{20mm}\times
  \bra{\xi_1',-\xi_1',\ldots,\xi_m',-\xi_m'}\sigma
  \ket{-\xi_n,\xi_n,\ldots,-\xi_1,\xi_1}\\*
  &\equiv&\sum_{m,n=0}^\infty C_{2m,2n}(t).
  \label{eq:Cmn}
\end{eqnarray}
Note that the indices $2m$ and $2n$ correspond to the number of
particles originating from the left and right initial state respectively.
Similarly, the expansion of the denominator reads
\begin{eqnarray}
  \bra{\Psi_0}\Psi_0\rangle&=&
  1+\sum_{n=1}^\infty\int_{0}^\infty
  \frac{d\xi_1'\ldots d\xi_n'}{n!(2\pi)^n}
  \frac{d\xi_1\ldots d\xi_n}{n!(2\pi)^n}
  \prod_{i=1}^n K_\text{q}(\xi_i')^*K_\text{q}(\xi_i)\,
  \bra{\xi_1',-\xi_1',\ldots,\xi_n',-\xi_n'}
  -\xi_n,\xi_n,\ldots,-\xi_1,\xi_1\rangle\quad\label{eq:BBa}\\*
  &\equiv&\sum_{n=0}^\infty Z_{2n}.
  \label{eq:BB}
\end{eqnarray}
The normalization in \eqref{eq:At} can therefore be formally expanded
in the following way
\begin{equation}
  \frac{1}{\bra{\Psi_0}\Psi_0\rangle}=
  1-Z_2+Z_2^2-Z_4+\ldots  
  \label{eq:1BB}
\end{equation}
We note that \eqref{eq:1BB} is merely used for defining linked
clusters, i.e. identifying the parts of the numerator in \eqref{eq:At}
that diverge in the infinite volume.

\subsection{Regularization procedure}\label{sec:reg}
The matrix elements in the terms $C_{2m,2n}(t)$ and $Z_{2n}$ possess kinematical 
poles whenever $\xi_i'=\xi_j$ and therefore have to be 
regularized. Following Smirnov~\cite{Smirnov92book} we proceed as follows:
Let $A$ denote a set of one-particle excitations and $A_1$ and $A_2$ a partition of $A$. 
The scattering matrix arising from the commutations
necessary to rewrite $\ket{A}$ as $\ket{A_2A_1}$ is denoted
by $S_{AA_1}$, i.e. $\ket{A}=S_{AA_1}\ket{A_2A_1}=S_{AA_2}\ket{A_1A_2}$. If $A$ and 
$B$ denote two sets of one-particle excitations, the form factors of $\sigma$ 
read~\cite{Smirnov92book,SE07}
\begin{eqnarray}
\bra{A}\sigma\ket{B}&=&
\sum_{\substack{A=A_1 \cup A_2\\ B=B_1 \cup B_2}}
d(B_2)\,S_{AA_1}\,S_{B_1B}\,\bra{A_2}B_2\rangle\,
\bra{A_1+\ii 0}\sigma\ket{B_1},\label{eq:reg}
\end{eqnarray}
where the sum is over all possible ways to break the sets $A=A_1\cup A_2$
and $B=B_1\cup B_2$ into subsets.  The scalar products $\bra{A_2}B_2\rangle$
as well as the corresponding terms in the $Z_{2n}$'s are easily evaluated using 
the Faddeev-Zamolodchikov algebra \eqref{eq:Aalgebra}. 
The factor $d(A)$ is present by virtue of the semi-locality of the 
spin operator with respect to the fundamental field and is given by
\begin{equation}
d(A)=(-1)^{n(A)},
\label{eq:defdA}
\end{equation}
where $n(A)$ denotes the number of elements in $A$. Using \eqref{eq:reg} the poles 
in the form factors have been shifted away from the real axis. The terms  
$\bra{A_2}B_2\rangle$ correspond to the disconnected pieces of the form factors.
As all rapidities in the remaining matrix elements are distinct, they can be evaluated 
using the crossing relation
\begin{equation}
\label{eq:crossing}
\bra{\theta_{1}'+\ii 0,\ldots,\theta_m'+\ii 0}
\sigma\ket{\theta_{1},\ldots,\theta_{n}}
=f(\theta_{1}'+\ii \pi+\ii \eta_{1},\ldots,
\theta_{m}'+\ii \pi+\ii \eta_{m},\theta_{1},\ldots,\theta_{n}),
\end{equation}
where $\eta_i\rightarrow 0^+$. We note that one can also
shift the rapidities in the set $A_1$ to the lower half plane, which
results~\cite{SE07} in different scattering and phase factors in
\eqref{eq:reg} but leaves the final result unchanged.  

It is clear that the right-hand side of \eqref{eq:reg} may still
contain divergences due to the intertwining of particles with
rapidities $\xi_i$ and $-\xi_i$ in the initial state
\eqref{eq:initialstate}. These divergences are a consequence of
working in the infinite volume and have to be canceled against similar
divergences originating from the norm of the initial state
\eqref{eq:BB}. In order to exhibit these cancellations we need to
identify these divergences explicitly. To this end we use the
$\kappa$-regularization scheme recently introduced in 
the study of finite-temperature correlation
functions~\cite{ek09}. For each pair of rapidities
$\{-\xi_i,\xi_i\}$ in the ket states we introduce an auxiliary real
parameter $\kappa_i$ to shift the rapidities away from the
singularities. The resulting expressions have to be understood as
generalized functions of the auxiliary variables $\kappa_i$. 
In order to exhibit the cancellations of terms in the Lehmann
representation of \eqref{eq:At} that diverge in the infinite volume
we define a smooth function $P(\kappa)$ which is strongly peaked
around $\kappa=0$ and satisfies 
\begin{equation}
P(0)=L,\quad \int d\kappa\,P(\kappa)=1.
\end{equation}
Here $L$ can be thought of as the length of the system in the
finite-volume regularization (see App.~\ref{app:finite}). One possible
choice is $P(\kappa)=L\,e^{-\pi L^2\kappa^2}$. 
Using this regularization scheme the first non-trivial term in the
expansion \eqref{eq:BB} reads
\begin{eqnarray}
Z_2&\equiv&
  \int d\kappa\,P(\kappa)\int_0^\infty\frac{d\xi'd\xi}{(2\pi)^2}\,K_\text{q}(\xi')^*K_\text{q}(\xi)\,
  \bra{\xi',-\xi'}-\xi+\kappa,\xi+\kappa\rangle\\
  &=&\int d\kappa\,P(\kappa)\,\delta(-2\kappa)\int_{\max\{0,-\kappa\}}^\infty d\xi\,
  K_\text{q}(\xi+\kappa)^*K_\text{q}(\xi)\nonumber\\*
  & &\qquad-\int d\kappa\,P(\kappa)\,\delta(-2\kappa)\int_0^{\max\{0,-\kappa\}} d\xi\,
  K_\text{q}(-\xi-\kappa)^*K_\text{q}(\xi)\label{eq:Z2calc}\\*
  &=&\frac{L}{2}\int_0^\infty d\xi\,\big|K_\text{q}(\xi)\big|^2.\label{eq:Z2calc2}
\end{eqnarray}
The infinite-volume divergence is now clearly exhibited and
\eqref{eq:Z2calc2} facilitates comparison with the finite-volume
regularization (see App.~\ref{app:finite}). Further examples for the
application of \eqref{eq:reg} and the $\kappa$-regularization are
presented in App.~\ref{app:BB}--\ref{app:D2m2m}. 

\subsection{Cancellation of singularities}\label{sec:cancel}
Using the $\kappa$-regularization as described in the previous section all terms in the 
expansions \eqref{eq:Cmn} and \eqref{eq:BB} are finite but may contain terms $\propto L^k$,
or equivalently $\propto\delta(-2\kappa_i)$,
that diverge in the infinite-volume limit. However, when we consider the one-point function
\eqref{eq:At} given by their quotient and expand again in powers of $K_\text{q}$,
\begin{equation}
\langle\sigma(t)\rangle=
\frac{\sum_{m,n=0}^\infty C_{2m,2n}(t)}{\sum_{n=0}^\infty Z_{2n}}\equiv
\sum_{m,n=0}^\infty D_{2m,2n}(t),
\label{eq:sigmaz}
\end{equation}
all terms $\propto L^k$ with $k\ge 1$ cancel each other and the remaining 
functions $D_{2m,2n}(t)$ are finite in the infinite-volume limit $L\to\infty$. 
This can be thought of as a linked-cluster expansion and is analogous
to the finite-temperature case \cite{ek09}.

\subsection{Resummation}
Performing the steps outlined in the proceeding sections we obtain the 
expansion \eqref{eq:sigmaz} for which the infinite-volume limit can safely be performed. 
After taking this limit we can study the long-time behaviour of the one-point function. 
Doing so we observe that the leading contribution to the term $D_{2m,2n}(t)$ will 
grow as $\propto t^\alpha$ with the power $\alpha$ depending on the number of particles
$2m$ and $2n$. As we will show below, however, these divergences can be 
resummed leading to a well-defined long-time behaviour of the one-point function,
which we present in Sec.~\ref{sec:longtime}.

\section{Results}\label{sec:results}
In this section we present the results for the leading terms in the expansion 
\eqref{eq:sigmaz}. We consider terms up to $\mathcal{O}(K_\text{q}^4)$ and
concentrate on the dominant contributions in the long-time limit. 
Technical details of the derivation are presented in Apps.~\ref{app:BB}--\ref{app:D44}.
In Sec.~\ref{sec:longtime} we present the final result for the long-time behaviour of
the one-point function \eqref{eq:At} after the resummation of the leading contributions
in the series \eqref{eq:sigmaz}.

\subsection{Terms in $\bs{\mathcal{O}(K_\text{q}^0)}$ and $\bs{\mathcal{O}(K_\text{q})}$}
The terms up to linear order in $K_\text{q}$ do not contain form factors possessing both 
incoming and outgoing particles. Thus there exist no kinematical poles, a regularization 
following the procedure discussed in Sec.~\ref{sec:reg} is not necessary and we
straightforwardly obtain
\begin{equation}
  D_{00}=C_{00}=\bra{0}\sigma\ket{0}=\bar{\sigma}
  \label{eq:D00}
\end{equation}
as well as
\begin{equation}
  D_{20}(t)+D_{02}(t)=C_{20}(t)+C_{02}(t)=
  \bar{\sigma}\int_0^\infty\frac{d\xi}{2\pi}\,\hat{K}_\text{q}(\xi)\,\tanh\xi
  \Bigl[e^{2M\ii t\cosh\xi}+e^{-2M\ii t\cosh\xi}\Bigr].
  \label{eq:D20}
\end{equation}
Here the real function $\hat{K}_\text{q}(\xi)$ was defined in \eqref{eq:quenchK}. The 
long-time behaviour of this term is obtained by a stationary phase approximation,
\begin{equation}
  D_{20}(t)+D_{02}(t)=-\frac{\bar{\sigma}}{8\sqrt{\pi}}\,
  \left(1-\frac{M}{M_0}\right)\,\frac{\cos(2Mt-\pi/4)}{(Mt)^{3/2}},\quad Mt\gg 1.
  \label{eq:D20longtime}
\end{equation}

\subsection{Terms in $\bs{\mathcal{O}(K_\text{q}^2)}$}
In this order there exist three terms. The first two originate from $C_{40}(t)$ and $C_{04}(t)$,
which do not possess kinematical poles. In the long-time limit we obtain
$D_{40}(t)+D_{04}(t)\sim\cos(4Mt)/(Mt)^5$ which 
constitutes a sub-leading correction to \eqref{eq:D20longtime}.

In contrast $C_{22}$ contains a form factor possessing both incoming and outgoing 
particles and hence kinematical poles appear. Performing the calculation as outlined in 
Secs.~\ref{sec:reg} and ~\ref{sec:cancel} we obtain (see App.~\ref{app:D22} for 
details of the derivation)
\begin{eqnarray}
    D_{22}(t)&=&C_{22}(t)-Z_2\bar{\sigma}=
    -\bar{\sigma}\Gamma\,t+D_{22}'(t),
    \label{eq:D22a}\\
    \Gamma&=&\frac{2M}{\pi}\,\int_0^\infty d\xi\,|K_\text{q}(\xi)|^2\,\sinh\xi,
    \label{eq:Gamma}\\
    D_{22}'(t)&=&\bar{\sigma}\,\mathfrak{Re}\,\int_0^\infty\frac{d\xi'}{2\pi}
  \int_{\gamma_-}\frac{d\xi}{2\pi}\,\hat{K}_\text{q}(\xi')\,\hat{K}_\text{q}(\xi)\,\tanh\xi'\,\tanh\xi\,
  \coth^2\frac{\xi'-\xi}{2}\,\coth^2\frac{\xi'+\xi}{2}\,e^{2M\ii t(\cosh\xi'-\cosh\xi)}.\label{eq:D22b}
  \end{eqnarray}
  Here the contour of integration $\gamma_-$ lies in the lower half plane and can be 
  explicitly parametrized by $(0<\phi_0\le\pi/4$)
\begin{equation}
\gamma_-(s)=\left\{\begin{array}{ll}-\ii s,& 0\le s\le \phi_0,\\
(s-\phi_0)-\ii\phi_0,& \phi_0\le s<\infty.
\end{array}\right.
\label{eq:contour}
\end{equation}  
Clearly the first term in \eqref{eq:D22a} dominates the long-time
behaviour. It can be thought of as the second term in the expansion of
$\bar{\sigma}\,e^{-\Gamma t}$ in powers of $K_\text{q}$, see
Sec.~\ref{sec:longtime}. The late-time behaviour of the second
contribution \eqref{eq:D22b} is dominated by the region $\xi'\approx
0$ and $\xi=-is$ with $s\approx 0$. Expanding the integrand and
changing to polar coordinates then gives
\be
D'_{22}(t)\approx -\frac{\bar{\sigma}}{32\pi}\frac{(1-M/M_0)^2}{Mt}
,\quad Mt\gg 1.
\label{eq:D22long}
\end{equation}
We note that the behaviour of \eqref{eq:D22a} is in agreement with the
finite-volume regularization presented in App.~\ref{app:finite}.

\subsection{Terms in $\bs{\mathcal{O}(K_\text{q}^3)}$}
In this order the terms containing kinematical poles are $C_{42}(t)$ and $C_{24}(t)$.
The calculation presented in App.~\ref{app:D42} yields
\begin{equation}
D_{42}(t)=D_{24}(t)^*=C_{42}(t)-Z_2\,C_{20}(t)=
-\Gamma\,t\,D_{20}(t)+\ldots,
\label{eq:D42}
\end{equation}
where the dots represent sub-leading terms that fall off at least as $\sim 1/(Mt)$ in the
long-time limit. 
Again we find a term showing an explicit linear time dependence, 
which can be viewed as the second term in the expansion of
$D_{20}(t)\,e^{-\Gamma t}$.

\subsection{Terms in $\bs{\mathcal{O}(K_\text{q}^4)}$}
The calculation of $D_{44}(t)$ requires the introduction of two independent
auxiliary parameters $\kappa_1$ and $\kappa_2$ (see App.~\ref{app:D44}). The result is
given by
\begin{eqnarray}
  D_{44}(t)&=&C_{44}(t)-Z_2C_{22}(t)+(Z_2^2-Z_4)\bar{\sigma}=
  \frac{\bar{\sigma}}{2}(\Gamma t)^2-\Gamma\,t\,D_{22}'(t)+D_{44}'+\ldots,\label{eq:D44a}\\*
  D_{44}'&=&\bar{\sigma}\,
  \mathfrak{Re}\,\int_0^\infty\frac{d\xi_1}{2\pi}\big|K_\text{q}(\xi_1)\big|^2\nonumber\\*
  & &\qquad\times\int_{\gamma_-}\frac{d\xi_2}{2\pi}\hat{K}_\text{q}(\xi_2)^2
  \left(\coth^2\frac{\xi_1+\xi_2}{2}-\tanh^2\frac{\xi_1+\xi_2}{2}-\coth^2\frac{\xi_1-\xi_2}{2}
  +\tanh^2\frac{\xi_1-\xi_2}{2}\right),\label{eq:D44b}
\end{eqnarray}
where the dots again represent sub-leading terms that fall off at least as $\sim 1/(Mt)$ in the
long-time limit. The first term in \eqref{eq:D44a} can be viewed as the third term in the 
expansion of $\bar{\sigma}\,e^{-\Gamma t}$, while the second corresponds to the second 
term in the expansion of $D_{22}'(t)\,e^{-\Gamma t}$. On the other hand $D_{44}'$ is 
independent of time and represents a correction of order $K_\text{q}^4$ to 
$D_{00}=\bar{\sigma}$. We further note that $D_{44}(t)$ does not contain terms that are 
linear in $t$ [as $D_{22}'(t)\propto 1/(Mt)$]. 

\subsection{Leading time dependence of higher-order terms}
Finally we argue in App.~\ref{app:D2m2m} that the leading term in the long-time behaviour of 
$D_{2m,2m}(t)$ and $D_{2m+2,2m}(t)$ are given by
\begin{eqnarray}
&&D_{2m,2m}(t)=\frac{\bar{\sigma}}{m!}(-\Gamma t)^m+\ldots,\label{eq:D2m2m}\\
&&D_{2m+2,2m}(t)=D_{2m,2m+2}(t)^*=
\frac{1}{m!}(-\Gamma t)^m\,D_{20}(t)+\ldots,\label{eq:D2m22m}
\end{eqnarray}
where the dots represent terms that grow at most as $\propto t^{m-1}$ for large times.

\subsection{Resummation and long-time behaviour}\label{sec:longtime}
As we have shown in the previous sections, the expansion \eqref{eq:sigmaz}
contains terms that grow at long times as powers of $t$.  
Hence, in order to obtain a well-defined result in the long-time limit
we have to resum these divergences. From the results presented in the 
previous sub-sections we deduce the leading long-time behaviour ($Mt\gg 1$) 
of the one-point function of $\sigma$ after a mass quench,
\begin{equation}
\langle\sigma(t)\rangle=
\bar{\sigma}\,\left[1+\frac{\alpha}{Mt}
-\frac{1-M/M_0}{8\sqrt{\pi}}\frac{\cos(2Mt-\pi/4)}{(Mt)^{3/2}}
+\ldots\right]
e^{-\Gamma t}+\ldots,
\label{eq:sigmat}
\end{equation}
where the dots represent sub-leading corrections to the prefactor as
well as terms that decay faster than $e^{-\Gamma t}$, respectively. 
The relaxation rate is given by
\begin{equation}
\Gamma(M,M_0)=
\frac{2M}{\pi}\int_0^\infty d\xi\,|K_\text{q}(\xi)|^2\,\sinh\xi+\mathcal{O}(K_\text{q}^6).
\label{eq:relaxationrate}
\end{equation}
We have determined the relaxation rate up to order
$K_\text{q}^6$. The fact that there is no contribution in
$\mathcal{O}(K_\text{q}^4)$ follows from the absence of terms linear
in $t$ in $D_{44}(t)$. This finding is in complete agreement
with the  corresponding result for the lattice model\cite{CEF-11}. 
The  consistent calculation of the relaxation rate in
$\mathcal{O}(K_\text{q}^6)$  would require the derivation of the
contributions to $D_{66}(t)$ that grow linearly in time. 

The $1$ and the $1/(Mt)^{3/2}$ term in the prefactor of the exponential in
\eqref{eq:sigmat} have been established by considering particular
contributions to all orders and showing that they exponentiate [see
\eqref{eq:D2m2m} and \eqref{eq:D2m22m}]. A detailed discussion of this point for
the lattice Ising chain is given in Ref.~[\onlinecite{CEF-12}]. On the
other hand,  the $\sim 1/(Mt)$ contribution in the prefactor is a
conjecture based on our results  for the leading contributions (in the
expansion in powers of the  quench matrix) in the Lehmann representation of
the one-point function. From \eqref{eq:D22long} and \eqref{eq:D44a} we
deduce 
\begin{equation}
\alpha=-\frac{(1-M/M_0)^2}{32\pi}+\mathcal{O}(K_\text{q}^3).
\end{equation}
Finally we stress that our results agree with the scaling limit 
of a quench in the transverse field of the Ising chain in the ordered 
phase\cite{CEF-11,CEF-12}.

\section{Extrapolation Time}\label{sec:FM}
The $K$-matrix for fixed boundary conditions in the Ising field theory is
given by\cite{GhoshalZamolodchikov94}
\begin{equation}
K_{\rm fixed}(\xi)=\ii\tanh\frac{\xi}{2}.
\end{equation}
We note that this is obtained as a limit of the quench $K$-matrix
(\ref{eq:quenchK})
\begin{equation}
K_{\rm fixed}(\xi)=\lim_{M_0\to\infty}K_q(\xi).
\end{equation}
In the quench problem a finite value of $M_0$ is required to render
rapidity integrals convergent at large energies. In particular, the
decay rate $\Gamma$ depends on $M_0$ and diverges in the 
limit $M_0\to\infty$. For quenches in interacting
integrable quantum field theories it is currently not known how to
express a given initial state in terms of eigenstates of
the post-quench Hamiltonian\cite{Spyros}. An exception are initial
states that correspond to integrable boundary conditions. In order to
use this information in the context of quantum quenches, a
prescription how to ``regularize'' the corresponding $K$-matrices at
large rapidities is required. Fioretto and Mussardo introduced an
``extrapolation time'' $\tau_0$ by the replacement\cite{fm-10}
\begin{equation}
K_{\rm fixed}(\xi)\to \ii\tanh\frac{\xi}{2}\,e^{-2M\tau_0\cosh\xi}
\equiv K_{\tau_0}(\xi).
\label{Kreg}
\end{equation}
Using this regularized $K$-matrix to perform our calculations results
in a decay rate
\be
\Gamma_{\tau_0}(M)=\frac{2M}{\pi}\int_0^\infty d\xi\,
|K_{\tau_0}(\xi)|^2\,\sinh\xi+\mathcal{O}(K_{\tau_0}^6).
\label{eq:relaxationrate_tau}
\ee
By comparing \eqref{eq:relaxationrate_tau} and
\eqref{eq:relaxationrate} and requiring the decay rates to be equal
$\Gamma_{\tau_0}(M)=\Gamma(M,M_0)$ is it possible to relate the
extrapolation time $\tau_0$ to the initial mass $M_0$.

\section{Conclusions}\label{sec:conclusion}
In this work we have considered the time evolution of the order
parameter after a quantum quench of the mass in the Ising field
theory. We have focussed on a quench within the ordered phase. We find
exponential decay of the order parameter to zero \eqref{eq:sigmat}.
Our results agree with the scaling limit of a quantum quench performed
in the ordered phase of the transverse field Ising chain\cite{CEF-11,CEF-12}. 
Our main achievement is of technical nature: we
have shown how to carry out calculations in the field theory
limit. Here, unlike for the lattice model, additional divergences
occur that need to be regulated appropriately. We have shown how to
use techniques developed recently in the study of integrable quantum
field theories at finite temperatures to overcome this problem. Our
method generalizes to \emph{interacting} integrable quantum field
theories such as the sine-Gordon and O($N$) non-linear sigma models. This
opens the door for analyzing quantum quenches in these theories, at least
for particular classes of initial states related to integrable
boundary conditions (``integrable quenches'')\cite{fm-10}. Work on the
sine-Gordon model is under way.

\section*{Acknowledgements}
We would like to thank P. Calabrese, M. Fagotti and D. Fioretto for
discussions. DS was supported by the German Research Foundation (DFG)
through the Emmy-Noether Program. FHLE acknowledges support by the
EPSRC under grant EP/I032487/1. 

\appendix
\section{Calculation of $\bra{\Psi_0}\Psi_0\rangle$}\label{app:BB}
In this appendix we evaluate the leading terms in the expansion 
\eqref{eq:BB} of the norm of the initial state $\ket{\Psi_0}$. Obviously one has
$Z_0=1$, while $Z_2$ was already calculated in Sec.~\ref{sec:reg} with the result
\begin{equation}
Z_2=\frac{L}{2}\int_0^\infty d\xi\,\big|K_\text{q}(\xi)\big|^2
\equiv \delta(-2\kappa)\int_0^\infty d\xi\,\big|K_\text{q}(\xi)\big|^2.
\label{eq:appZ2}
\end{equation}
In the last step we have reintroduced the auxiliary variable $\kappa$. We will
explicitly retain the auxiliary variables $\kappa_i$ throughout the appendices, but
keep in mind that all expressions have to be understood as generalized functions 
of $\kappa_i$ as discussed in Sec.~\ref{sec:reg}. As all expressions are multiplied
by the strongly peaked functions $P(\kappa_i)$ we can drop all terms $\propto\kappa_i^n$
with $n\ge 1$. In contrast, all irregular terms $\propto\delta(\kappa_i)$ as well as all
divergent terms $\propto\kappa_1^n$ with $n\le -1$ have to cancel when considering 
the expansion \eqref{eq:sigmaz}. It is the purpose of these appendices to show 
by explicit evaluation up to $\mathcal{O}(K_\text{q}^4)$ that these terms indeed 
cancel each other and that the remaining terms $\propto\kappa_i^0$ yield the results
for $D_{2m,2n}(t)$ presented in Sec.~\ref{sec:results}.

The next term $Z_4$ requires the introduction of two auxiliary variables
$\kappa_1$ and $\kappa_2$. Starting from \eqref{eq:BBa} we have to regularize 
the overlap element
\begin{eqnarray}
&&\bra{\xi_1',-\xi_1',\xi_2',-\xi_2'}-\xi_2,\xi_2,-\xi_1,\xi_1\rangle\equiv
\bra{\xi_1',-\xi_1',\xi_2',-\xi_2'}
-\xi_2+\kappa_2,\xi_2+\kappa_2,-\xi_1+\kappa_1,\xi_1+\kappa_1\rangle\\[2mm]
&&\qquad\qquad=(2\pi)^4\Bigl[
\delta(\xi_1'-\xi_1-\kappa_1)\delta(-\xi_1'+\xi_1-\kappa_1)
\delta(\xi_2'-\xi_2-\kappa_2)\delta(-\xi_2'+\xi_2-\kappa_2)\nonumber\\*
& &\qquad\qquad\phantom{=(2\pi)^4}
-\delta(\xi_1'-\xi_1-\kappa_1)\delta(-\xi_1'+\xi_2-\kappa_2)
\delta(\xi_2'-\xi_2-\kappa_2)\delta(-\xi_2'+\xi_1-\kappa_1)\nonumber\\*
& &\qquad\qquad\phantom{=(2\pi)^4}
-\delta(\xi_1'-\xi_2-\kappa_2)\delta(-\xi_1'+\xi_1-\kappa_1)
\delta(\xi_2'-\xi_1-\kappa_1)\delta(-\xi_2'+\xi_2-\kappa_2)\nonumber\\*
& &\qquad\qquad\phantom{=(2\pi)^4}
+\delta(\xi_1'-\xi_2-\kappa_2)\delta(-\xi_1'+\xi_2-\kappa_2)
\delta(\xi_2'-\xi_1-\kappa_1)\delta(-\xi_2'+\xi_1-\kappa_1)+\ldots\Bigr].\label{eq:Z48}
\end{eqnarray}
Here the dots represent 12 further combinations of $\delta$-functions which lead, 
in analogy to \eqref{eq:Z2calc}, to terms containing integrals restricted to intervals like 
$0<\xi_i<\kappa_i$. These terms in turn yield contributions $\propto\kappa_i^n$ with
$n\ge 1$ which vanish when performing the $\kappa$-integrations. Hence we have
not written these terms in \eqref{eq:Z48}. Now straightforward evaluation of the four 
terms yields
\begin{equation}
  Z_4=
  \frac{1}{2}\delta(-2\kappa_1)\delta(-2\kappa_2)
  \left(\int_{0}^\infty d\xi\,\big|K_\text{q}(\xi)\big|^2\right)^2
  -\frac{1}{2}\delta(-2\kappa_1-2\kappa_2)\int_{0}^\infty d\xi\,K_\text{q}(\xi+\kappa_1)^*\,
  K_\text{q}(\xi-\kappa_1)^*\,K_\text{q}(\xi)^2.
  \label{eq:appZ4}
\end{equation}

\section{Calculation of $\bs{D_{22}}$}\label{app:D22}
The first term in the expansion \eqref{eq:Cmn} which involves kinematical poles is 
$C_{22}(t)$, which after shifting the rapidities in the ket by $\kappa$ reads
\begin{equation}
    C_{22}(t)=\int_{0}^\infty\frac{d\xi'd\xi}{(2\pi)^2}\,K_\text{q}(\xi')^*\,K_\text{q}(\xi)\,
    \bra{\xi',-\xi'}\sigma\ket{-\xi+\kappa,\xi+\kappa}\,
    e^{2M\ii t(\cosh\xi'-\cosh\xi)}.
    \label{eq:C22}
\end{equation}
We decompose the form factor into its connected and disconnected pieces using
\eqref{eq:reg}    
\begin{eqnarray}
    \bra{\xi',-\xi'}\sigma\ket{-\xi+\kappa,\xi+\kappa}&=&
    (2\pi)^2\,\bar{\sigma}\,
    \bigl[\delta(\xi'-\xi-\kappa)\,\delta(-\xi'+\xi-\kappa)
    -\delta(\xi'+\xi-\kappa)\,\delta(-\xi'-\xi-\kappa)\bigr]\nonumber\\*
    & &-2\pi\,\delta(\xi'-\xi-\kappa)\,f(-\xi'+\ii\pi+\ii\eta,-\xi+\kappa)\nonumber\\*
    & &-2\pi\,\delta(-\xi'+\xi-\kappa)\,f(\xi'+\ii\pi+\ii\eta,\xi+\kappa)\nonumber\\*
    & &+2\pi\,\delta(\xi'+\xi-\kappa)\,f(-\xi'+\ii\pi+\ii\eta,\xi+\kappa)\nonumber\\*
    & &+2\pi\,\delta(-\xi'-\xi-\kappa)\,f(\xi'+\ii\pi+\ii\eta,-\xi+\kappa)\nonumber\\*
    & &+f(\xi'+\ii\pi+\ii\eta_1,-\xi'+\ii\pi+\ii\eta_2,-\xi+\kappa,\xi+\kappa)
    \label{eq:reg1}  
\end{eqnarray}
with $\eta,\eta_i\rightarrow 0^+$. Insertion of \eqref{eq:reg1} into \eqref{eq:C22} 
yields three different types of terms, which we denote by $C_{22}^0$, $C_{22}^1(t)$, 
and $C_{22}^2(t)$ respectively. 

The first line simply gives
\begin{equation}
    C_{22}^0=\bar{\sigma}\,\delta(-2\kappa)\,
    \int_{0}^\infty d\xi\,\big|K_\text{q}(\xi)\big|^2.
\end{equation}
  
The second term is obtained from the second to fifth lines in \eqref{eq:reg1}, which yield
\begin{equation}
	C_{22}^1(t)=\ii\bar{\sigma}\,\coth\frac{2\kappa-\ii\eta}{2}
	\int_{-\infty}^\infty\frac{d\xi}{2\pi}\,K_\text{q}(\xi+\kappa)^*\,K_\text{q}(\xi)\,
	e^{2M\ii t[\cosh(\xi+\kappa)-\cosh\xi]}.
	\label{eq:C221}
\end{equation}
Hereby we have already omitted terms of the form
\begin{equation}
  	-\ii\bar{\sigma}\,\coth\frac{2\kappa-\ii\eta}{2}
	\underbrace{\int_0^\kappa\frac{d\xi}{2\pi}\,K_\text{q}(\kappa-\xi)^*\,K_\text{q}(\xi)\,
	e^{2M\ii t[\cosh(\kappa-\xi)-\cosh\xi]}}_{\approx |K_\text{q}(\kappa/2)|^2\,\kappa}
	\propto\kappa^2
\end{equation}
which vanishes in the $\kappa$-regularization scheme due to
\begin{equation*}
\int d\kappa\,P(\kappa)\,\kappa^n\to 0\quad\text{for}\quad L\to\infty,\;n\ge 1.
\end{equation*}
We can further analyze \eqref{eq:C221} by expanding the integrand up to 
$\mathcal{O}(\kappa)$ 
\begin{equation}
K_\text{q}(\xi+\kappa)^*\,K_\text{q}(\xi)\,e^{2M\ii t[\cosh(\xi+\kappa)-\cosh\xi]}
=\big|K_\text{q}(\xi)\big|^2
+\kappa\hat{K}_\text{q}(\xi)\,\frac{d\hat{K}_\text{q}}{d\xi}(\xi)
+2M\ii t\kappa\big|K_\text{q}(\xi)\big|^2\,\sinh\xi+\mathcal{O}(\kappa^2).
\end{equation}
The contributions from the second and third term vanish as they are antisymmetric 
under $\xi\to-\xi$, thus we arrive at
\begin{equation}
C_{22}^1(t)=2\ii\bar{\sigma}\,\coth\frac{2\kappa-\ii\eta}{2}
	\int_0^\infty\frac{d\xi}{2\pi}\,|K_\text{q}(\xi)|^2.
	\label{eq:appC221a}
\end{equation}

Finally, the sixth line yields 
\begin{eqnarray}
  C_{22}^2(t)&=&\bar{\sigma}\int_0^\infty\frac{d\xi'd\xi}{(2\pi)^2}\,K_\text{q}(\xi')^*\,
  K_\text{q}(\xi)\,
  \tanh\xi'\,\tanh\xi\,e^{2M\ii t(\cosh\xi'-\cosh\xi)}\qquad\nonumber\\*
  & &\times\coth\frac{\xi'+\xi-\kappa+\ii\eta_1}{2}\,
  \coth\frac{\xi'-\xi-\kappa+\ii\eta_1}{2}\,\coth\frac{\xi'-\xi+\kappa-\ii\eta_2}{2}\,
  \coth\frac{\xi'+\xi+\kappa-\ii\eta_2}{2}.
  \label{eq:C222}
\end{eqnarray}
In order to isolate the singularities we may shift the $\xi'$-contour to the upper half plane or
the $\xi$-contour to the lower half plane. Doing so we pick up contributions from the poles at 
$\xi'=\xi-\kappa+\ii\eta_{2}$, $\xi'=-\xi-\kappa+\ii\eta_{2}$ and 
$\xi=\xi'+\kappa-\ii\eta_{2}$, $\xi=-\xi'+\kappa-\ii\eta_{1}$, respectively, and 
we obtain
\begin{eqnarray}
  C_{22}^2(t)&=&\bar{\sigma}\,\mathfrak{Re}\,\int_0^\infty\frac{d\xi'}{2\pi}
  \int_{\gamma_-}\frac{d\xi}{2\pi}\,\hat{K}_\text{q}(\xi')\,\hat{K}_\text{q}(\xi)\,\tanh\xi'\,\tanh\xi\,
  \coth^2\frac{\xi'-\xi}{2}\,\coth^2\frac{\xi'+\xi}{2}\,e^{2M\ii t(\cosh\xi'-\cosh\xi)}
  \qquad\label{eq:C222a}\\
  & &-\ii\bar{\sigma}\coth\frac{2\kappa-\ii(\eta_1+\eta_2)}{2}
  \int_0^\infty\frac{d\xi}{2\pi}\,\hat{K}_\text{q}(\xi)\,e^{-2M\ii t\cosh\xi}\nonumber\\*
  & &\hspace{20mm}\times\left\{
  \Theta(\xi-\kappa)\,\hat{K}_\text{q}(\xi-\kappa)\,e^{2M\ii t\cosh(\xi-\kappa)}+
  \Theta(-\xi-\kappa)\,\hat{K}_\text{q}(-\xi-\kappa)\,e^{2M\ii t\cosh(\xi+\kappa)}\right\}
  \label{eq:C222b}\\
  & &-\ii\bar{\sigma}\coth\frac{2\kappa-\ii(\eta_1+\eta_2)}{2}
  \int_0^\infty\frac{d\xi}{2\pi}\,\hat{K}_\text{q}(\xi)\,e^{2M\ii t\cosh\xi}\nonumber\\*
  & &\hspace{20mm}\times\left\{
  \Theta(\xi+\kappa)\,\hat{K}_\text{q}(\xi+\kappa)\,e^{-2M\ii t\cosh(\xi+\kappa)}-
  \Theta(-\xi+\kappa)\,\hat{K}_\text{q}(-\xi+\kappa)\,e^{-2M\ii t\cosh(\xi-\kappa)}
  \right\},\qquad\label{eq:C222c}
\end{eqnarray}
where the path $\gamma_-$ lies in the lower half plane and was explicitly 
defined in \eqref{eq:contour}. Expanding \eqref{eq:C222b} and \eqref{eq:C222c}
in $\kappa$ yields
\begin{equation}
-2\ii\bar{\sigma}\coth\frac{2\kappa-\ii(\eta_1+\eta_2)}{2}
  \int_0^\infty\frac{d\xi}{2\pi}\,|K_\text{q}(\xi)|^2-
  \frac{2M\bar{\sigma} t}{\pi}\,\kappa\,\coth\frac{2\kappa-\ii(\eta_1+\eta_2)}{2}
  \int_0^\infty d\xi\,|K_\text{q}(\xi)|^2\,\sinh\xi,\label{eq:C222d}
\end{equation}
where we have used
\begin{equation}
\int_0^\infty d\xi\,\hat{K}_\text{q}(\xi)\,\frac{d\hat{K}_\text{q}}{d\xi}(\xi)=\frac{1}{2}\hat{K}_\text{q}(\xi)^2\Big|_0^\infty=0.
\end{equation}
Now the first term in \eqref{eq:C222d} cancels $C_{22}^1(t)$ and we arrive at the final result
\begin{equation}
C_{22}(t)=\bar{\sigma}\,\delta(-2\kappa)\,
    \int_{0}^\infty d\xi\,\big|K_\text{q}(\xi)\big|^2
    -\frac{2M\bar{\sigma} t}{\pi}\,\int_0^\infty d\xi\,|K_\text{q}(\xi)|^2\,\sinh\xi+D_{22}'(t)
    =\bar{\sigma} Z_2-\bar{\sigma}\Gamma t+D_{22}'(t),
    \label{eq:appC22}
\end{equation}
where $Z_2$, $\Gamma$ and $D_{22}'(t)$ are defined by \eqref{eq:appZ2}, 
\eqref{eq:Gamma}, and \eqref{eq:C222a} or \eqref{eq:D22b} respectively. 
We stress that the first term exactly cancels the product $Z_2C_{00}=\bar{\sigma} Z_2$, while
the second term equals the second term in the expansion of $\bar{\sigma}\,e^{-\Gamma t}$.

\section{Calculation of $\bs{D_{42}}$ and $\bs{D_{24}}$}\label{app:D42}
We restrict ourselves to $C_{42}(t)$ as $C_{24}(t)=C_{42}(t)^*$. For the calculation of 
$C_{42}(t)$ we follow the same steps as above: (i) shift the rapidities in the ket by the auxiliary 
variable $\ket{-\xi,\xi}\to\ket{-\xi+\kappa,\xi+\kappa}$, (ii) analytically continue the resulting
form factor using \eqref{eq:reg}, (iii) evaluate the terms by shifting the contours of integration
for $\xi_{1}'$ and $\xi_2'$ to the upper half plane, and (iv) expand the result in $\kappa$ 
up to $\mathcal{O}(\kappa^0)$. In particular one can show that all ill-defined terms 
$\propto 1/\kappa$ [see \eqref{eq:appC221a} for a similar term in $\mathcal{O}(K_\text{q}^2)$]
cancel each other. After straightforward calculation we obtain
\begin{eqnarray}
C_{42}(t)&=&\delta(-2\kappa)\,C_{20}(t)\,\int_0^\infty\,d\xi\,\big|K_\text{q}(\xi)\big|^2
-\Gamma\,t\,C_{20}(t)
+\bar{\sigma}\int_0^\infty\frac{d\xi}{2\pi}\hat{K}_\text{q}(\xi)^3\,\tanh\xi\,e^{2M\ii t\cosh\xi}
\label{eq:appC42}\\
& &-2\ii\bar{\sigma}\int_{\gamma_+}\frac{d\xi'}{2\pi}\int_0^\infty\frac{d\xi}{2\pi}\,
  \hat{K}_\text{q}(\xi')\,\big|K_\text{q}(\xi)\big|^2\,\tanh\xi'\nonumber\\*
 & &\qquad\times \left(\coth\frac{\xi'-\xi}{2}-\tanh\frac{\xi'-\xi}{2}-\coth\frac{\xi'+\xi}{2}
  +\tanh\frac{\xi'+\xi}{2}\right)e^{2M\ii t\cosh\xi'}\\
&&+\frac{\bar{\sigma}}{2}\int_{\gamma_+}\frac{d\xi_1'd\xi_2'}{(2\pi)^2}
\int_0^\infty\frac{d\xi}{2\pi}\,\hat{K}_\text{q}(\xi)\,\tanh\xi\,
\prod_{i=1}^2\hat{K}_\text{q}(\xi_i')\,\tanh\xi_i'\,
\prod_{i=1}^2\coth^2\frac{\xi_i'-\xi}{2}\,\coth^2\frac{\xi_i'+\xi}{2}\nonumber\\*
& &\qquad\qquad\qquad\qquad\qquad
\times\tanh^2\frac{\xi_1'-\xi_2'}{2}\,\tanh^2\frac{\xi_1'+\xi_2'}{2}\,
e^{2M\ii t(\cosh\xi_1'+\cosh\xi_2'-\cosh\xi)},
\end{eqnarray}
where the path $\gamma_+$ lies in the upper half plane and is explicitly 
defined by $(0<\phi_0\le\pi/4$)
\begin{equation}
\gamma_+(s)=\left\{\begin{array}{ll}\ii s,& 0\le s\le \phi_0,\\
(s-\phi_0)+\ii\phi_0,& \phi_0\le s<\infty.
\end{array}\right.
\label{eq:contour2}
\end{equation}
The first term in \eqref{eq:appC42} cancels against the product of $Z_2\,C_{20}(t)$,
while the second term corresponds to the second term in the expansion of 
$C_{20}(t)\,e^{-\Gamma t}$. All other terms constitute sub-leading corrections to 
\eqref{eq:sigmat}.

\section{Calculation of $\bs{D_{44}}$}\label{app:D44}
The calculation follows the same steps as outlined above. The main difference is that
we have to introduce two auxiliary variables, i.e. the form factor in $C_{44}(t)$ becomes
\begin{equation}
\bra{\xi_1',-\xi_1',\xi_2',-\xi_2'}\sigma\ket{-\xi_2,\xi_2,-\xi_1,\xi_1}\to
\bra{\xi_1',-\xi_1',\xi_2',-\xi_2'}\sigma
\ket{-\xi_2+\kappa_2,\xi_2+\kappa_2,-\xi_1+\kappa_1,\xi_1+\kappa_1},
\end{equation}
which is analytically continued using \eqref{eq:reg}. After a tedious but straightforward 
evaluation of the resulting terms up to $\mathcal{O}(\kappa_1^0,\kappa_2^0)$ one can 
explicitly show that all ill-defined terms
\begin{equation}
\propto\coth\frac{2\kappa_1-\ii 0}{2},\;\propto\coth\frac{2\kappa_2-\ii 0}{2},\;
\propto\coth\frac{2\kappa_1+2\kappa_2-\ii 0}{2}
\end{equation}
cancel each other, where we have employed for example
\begin{equation}
\int d\kappa_1\,d\kappa_2\,\kappa_1^m\,\kappa_2^n\,\coth\frac{2\kappa_2-\ii 0}{2}\,
P(\kappa_1)\,P(\kappa_2)\propto L^{1-m-n}\to 0\quad\text{for}\quad m+n\ge 2.
\end{equation}
The final result is
\begin{eqnarray}
&&C_{44}(t)=\bar{\sigma}\,Z_4
  +\frac{1}{2}\Bigl(\delta(-2\kappa_1)+\delta(-2\kappa_2)\Bigr)\,
  \Bigl(D_{22}'(t)-\bar{\sigma}\Gamma t\Bigr)\int_0^\infty d\xi\,\big|K_\text{q}(\xi)\big|^2
  +\frac{\bar{\sigma}}{2}(\Gamma t)^2-\Gamma\,t\,D_{22}'(t)\label{eq:C44a}\\
  & &\qquad+\bar{\sigma}\,\mathfrak{Re}\,
  \int_0^\infty\frac{d\xi_1}{2\pi}\big|K_\text{q}(\xi_1)\big|^2\nonumber\\*
  & &\qquad\qquad\qquad\times\,\int_{\gamma_-}\frac{d\xi_2}{2\pi}\hat{K}_\text{q}(\xi_2)^2
  \left(\coth^2\frac{\xi_1+\xi_2}{2}-\tanh^2\frac{\xi_1+\xi_2}{2}-\coth^2\frac{\xi_1-\xi_2}{2}
  +\tanh^2\frac{\xi_1-\xi_2}{2}\right)\qquad\label{eq:C44b}\\
  & &\qquad+2\bar{\sigma}\,\mathfrak{Re}\,\int_0^\infty\frac{d\xi'}{2\pi}
  \int_{\gamma_-}\frac{d\xi}{2\pi}\,\hat{K}_\text{q}(\xi')^3\,\hat{K}_\text{q}(\xi)\,\tanh\xi'\,\tanh\xi\,
  \coth^2\frac{\xi'-\xi}{2}\,\coth^2\frac{\xi'+\xi}{2}\,e^{2M\ii t(\cosh\xi'-\cosh\xi)}\label{eq:C44c}\\
  & &\qquad+2\bar{\sigma}\, \int_{\gamma_+}\frac{d\xi'}{2\pi}\,\int_0^\infty\frac{d\xi_1}{2\pi}
  \int_{\gamma_-}\frac{d\xi_2}{2\pi}\,\hat{K}_\text{q}(\xi')\,\hat{K}_\text{q}(\xi_1)^2\,
  \hat{K}_\text{q}(\xi_2)\,\tanh\xi'\,\tanh\xi_2\,
  \coth^2\frac{\xi'-\xi_2}{2}\,\coth^2\frac{\xi'+\xi_2}{2}\nonumber\\*
  & &\qquad\qquad\qquad\times\,\left[\tanh\frac{\xi'-\xi_1}{2}-\tanh\frac{\xi'+\xi_1}{2}-
  \coth\frac{\xi'-\xi_1}{2}+\coth\frac{\xi'+\xi_1}{2}\right.\nonumber\\*
  & &\qquad\qquad\qquad\qquad\left.+\tanh\frac{\xi_1-\xi_2}{2}-\tanh\frac{\xi_1+\xi_2}{2}-
  \coth\frac{\xi_1-\xi_2}{2}+\coth\frac{\xi_1+\xi_2}{2}
  \right]e^{2M\ii t(\cosh\xi'-\cosh\xi_2)}\label{eq:C44d}\\
  & &\qquad+\frac{\bar{\sigma}}{4} \int_{\gamma_+}\frac{d\xi_1'd\xi_2'}{(2\pi)^2}\,
  \int_0^\infty\frac{d\xi_1d\xi_2}{(2\pi)^2}\,
  \prod_{i=1}^2\hat{K}_\text{q}(\xi_i')\,\hat{K}_\text{q}(\xi_i)\,\tanh\xi_i'\,\tanh\xi_i\,
  \prod_{i,j=1}^2\coth^2\frac{\xi_i'-\xi_j}{2}\,\coth^2\frac{\xi_i'+\xi_j}{2}\nonumber\\*
  & &\qquad\qquad\qquad\qquad\times\tanh^2\frac{\xi_1-\xi_2}{2}\,\tanh^2\frac{\xi_1+\xi_2}{2}\,
  \tanh^2\frac{\xi_1'-\xi_2'}{2}\,\tanh^2\frac{\xi_1'+\xi_2'}{2}\,
  e^{2M\ii t\sum_i(\cosh\xi_i'-\cosh\xi_i)},\label{eq:C44e}
\end{eqnarray}
where $Z_4$, $\Gamma$ and $D_{22}'(t)$ are defined in \eqref{eq:appZ4},  
\eqref{eq:Gamma} and \eqref{eq:C222a} respectively. The paths $\gamma_\pm$ are
defined in \eqref{eq:contour} and \eqref{eq:contour2}. The leading contributions are given 
by the first line \eqref{eq:C44a}, the second line \eqref{eq:C44b} yields the time-independent
term $D_{44}'$, and \eqref{eq:C44c}--\eqref{eq:C44e} constitute sub-leading corrections
that fall off at least as $\sim 1/(Mt)$ in the long-time limit.

Now $D_{44}(t)$ is obtained by [see \eqref{eq:appC22}]
\begin{eqnarray}
D_{44}(t)&=&C_{44}(t)-Z_2C_{22}(t)+(Z_2^2-Z_4)\bar{\sigma}\\
&=&\left(\delta(-2\kappa)-\frac{1}{2}\delta(-2\kappa_1)-\frac{1}{2}\delta(-2\kappa_2)\right)
\bigl(\bar{\sigma}\Gamma t-D_{22}'(t)\bigr)\int_0^\infty d\xi\,\big|K_\text{q}(\xi)\big|^2
\label{eq:appD44a}\\*
&&+\frac{\bar{\sigma}}{2}(\Gamma t)^2-\Gamma\,t\,D_{22}'(t)+D_{44}'+\ldots,
\end{eqnarray}
where the dots represent the sub-leading terms \eqref{eq:C44c}--\eqref{eq:C44e}.
Here \eqref{eq:appD44a} vanishes due to 
\begin{eqnarray}
\delta(-2\kappa)-\frac{1}{2}\delta(-2\kappa_1)-\frac{1}{2}\delta(-2\kappa_2)
&\equiv&\int d\kappa\,\delta(-2\kappa)\,P(\kappa)
-\frac{1}{2}\int d\kappa_1\,d\kappa_2\,\bigl[\delta(-2\kappa_1)+\delta(-2\kappa_2)\bigr]
P(\kappa_1)\,P(\kappa_2)\\
&=&\frac{L}{2}-\frac{L}{4}\int d\kappa_2\,P(\kappa_2)-\frac{L}{4}\int d\kappa_1\,P(\kappa_1)
=0,
\end{eqnarray}
and we arrive at \eqref{eq:D44a}.

\section{Leading time dependence of $\bs{D_{2m,2m}}$ and $\bs{D_{2m+2,2m}}$}
\label{app:D2m2m}
The leading behaviour of $D_{2m,2m}(t)$ in the long-time limit is obtained from 
$C_{2m,2m}(t)$ given in \eqref{eq:Cmn} by (i) introducing the auxiliary variables 
$\kappa_i$, $i=1,\ldots,m$, (ii) regularizing the form factor according to \eqref{eq:reg}
while keeping only the connected piece (i.e. the term with $A_2=B_2=\emptyset$),
(iii) evaluating the connected form factor \eqref{eq:sigmaff}, and (iv) shifting
the $\xi_i'$-contours to the upper half plane and keeping only the contributions from the
poles at $\xi_i'=\xi_j-\kappa_j+\ii 0$. The result after these steps reads
\begin{equation}
\begin{split}
&\bar{\sigma}\frac{(-2\ii)^m}{m!}\prod_{i=1}^m\coth\frac{2\kappa_i-\ii 0}{2}
\int_0^\infty\frac{d\xi_1\ldots d\xi_m}{(2\pi)^m}
\prod_{i=1}^mK_\text{q}(\xi_i-\kappa_i)^*\,K_\text{q}(\xi_i)\,
e^{2M\ii t(\cosh(\xi-\kappa_i)-\cosh\xi_i)}\\
&\quad\times\prod_{\substack{i,j=1\\ i<j}}^{m}
\tanh\frac{\xi_i-\xi_j+\kappa_i-\kappa_j}{2}\,\tanh\frac{\xi_i-\xi_j-\kappa_i+\kappa_j}{2}\,
\coth\frac{\xi_i-\xi_j-\kappa_i-\kappa_j+\ii 0}{2}\,
\coth\frac{\xi_i-\xi_j+\kappa_i+\kappa_j-\ii 0}{2}.
\end{split}
\end{equation}
Now expanding in the $\kappa_i$'s we obtain
\begin{equation}
\bar{\sigma}\frac{(-4Mt)^m}{m!}\int_0^\infty\frac{d\xi_1\ldots d\xi_m}{(2\pi)^m}
\prod_{i=1}^m\big|K_\text{q}(\xi_i)\big|^2\,\sinh\xi_i+\ldots
=\frac{\bar{\sigma}}{m!}\left(-\frac{2M}{\pi}\int_0^\infty d\xi\,
\big|K_\text{q}(\xi)\big|^2\,\sinh\xi\right)^m t^m+\ldots,
\end{equation}
where the dots represent terms that grow at most as $\propto t^{m-1}$ for large times.
As the disconnected pieces of $C_{2m,2m}(t)$ also grow at most as $\propto t^{m-1}$
we deduce that the leading time dependence of $D_{2m,2m}(t)$ is given 
by \eqref{eq:D2m2m}. Following the same line of argument for $C_{2m+2,2m}(t)$ 
we arrive at \eqref{eq:D2m22m}.

\section{Finite-size regularization}\label{app:finite}
In our calculation in the infinite-volume system we have regularized the kinematical 
poles in the form factors using a combination of the analytic continuation \eqref{eq:reg}
together with the $\kappa$-regularization. An alternative procedure to regularize the
kinematical poles is to study~\cite{KormosPozsgay10,CEF-11,CEF-12} the 
system with a finite length $L$. In this case the Hilbert space divides itself into two
sectors: the Neveu--Schwarz (NS) sector corresponding to antiperiodic boundary conditions
and the Ramond (R) sector corresponding to periodic ones. The rapidities in these
sectors are quantized according to 
\begin{eqnarray}
\text{NS}:& &ML\sinh\xi_p=2\pi p,\; p\in\mathbb{Z}+\tfrac{1}{2},\\
\text{R}:& &ML\sinh\theta_q=2\pi q,\; q\in\mathbb{Z}.
\end{eqnarray}
The spin operator connects the two sectors, its form factors in the
finite system read~\cite{Bugrii} 
\begin{equation}
\label{eq:finiteff}
_\mathrm{NS}\!\bra{p_1,\ldots,p_m}\sigma
\ket{q_1,\ldots,q_n}_\mathrm{R}=
S(L)\prod_{i=1}^m\tilde{g}(\xi_{p_i})\prod_{j=1}^ng(\theta_{q_j})
F_{mn}(\xi_{p_1},\ldots,\xi_{p_m}|\theta_{q_1},\ldots,\theta_{q_n}),
\end{equation}
where the function $F_{mn}$ is the infinite-volume form factor 
[see \eqref{eq:sigmaff} and \eqref{eq:crossing}]
\begin{equation}
F_{mn}(\xi_{1},\ldots,\xi_{m}|\theta_{1},\ldots,\theta_{n})=
\ii^{\lfloor(m+n)/2\rfloor}\bar{\sigma} 
\prod_{\substack{i,j=1\\ i<j}}^m\tanh\frac{\xi_i-\xi_j}{2}
\prod_{\substack{i,j=1\\ i<j}}^n\tanh\frac{\theta_i-\theta_j}{2}
\prod_{i=1}^m\prod_{j=1}^n\coth\frac{\xi_i-\theta_j}{2},
\label{eq:finiteF}
\end{equation}
the symbol $\lfloor \; \rfloor$ denotes the floor function, and the constant as well as the leg
factors are
\begin{equation}
S(L)=1+\mathcal{O}(e^{-L}),\quad
g(\theta)=\tilde{g}(\theta)=\frac{1}{\sqrt{ML\cosh\theta}}+\mathcal{O}(e^{-L}).
\end{equation}
We stress that due to the quantization of the rapidities the singularities in the 
form factor \eqref{eq:finiteF} are regularized.

\begin{figure}[t]
\begin{center}
\includegraphics[width=75mm]{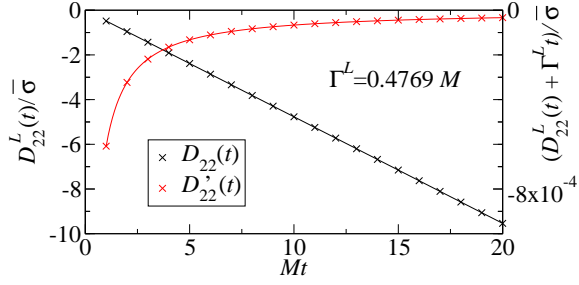}
\caption{$D_{22}^L(t)$ (black line, left axis) and $D_{22}^L(t)+\Gamma^Lt$ 
(red line, right axis) for $\tau_0=0.1$, $L=50$ and $N=300$. For comparison we 
show $D_{22}(t)$ (black stars) and  $D_{22}'(t)$ (red stars) as defined in \eqref{eq:D22a} 
and \eqref{eq:D22b} respectively. We observe excellent agreement between the 
finite-volume regularization and the infinite-volume results.}
\label{fig:fvreg1}
\end{center}
\end{figure}

\begin{figure}[t]
\begin{center}
\includegraphics[width=75mm]{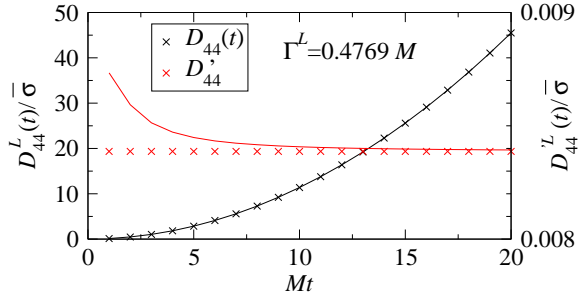}
\caption{$D_{44}^L(t)$ (black line, left axis) and 
$D_{44}^{'L}(t)\equiv D_{44}^L(t)-(\Gamma^Lt)^2/2+\Gamma^Lt(D_{22}^L(t)+\Gamma^Lt)$ 
(red line, right axis) for $\tau_0=0.1$, $L=50$ and $N=300$. For comparison we 
show $D_{44}(t)$ (black stars) and  $D_{44}'$ (red stars) as defined in \eqref{eq:D44a} 
and \eqref{eq:D44b} respectively. Note that in contrast to $D_{44}'$ the finite-volume term 
$D_{44}^{'L}(t)$ depends on the time as it also includes corrections which fall off as powers
of $Mt$ [these corrections are incorporated in the dots in \eqref{eq:D44a}]. 
We observe excellent agreement between the 
finite-volume regularization and the infinite-volume results.}
\label{fig:fvreg2}
\end{center}
\end{figure}

As we consider quenches in the ordered phase which breaks the $\mathbb{Z}_2$ 
invariance the initial state in a system of length $L$ has the form~\cite{CEF-11,CEF-12}
\begin{equation}
\ket{\Psi_0}_L=\frac{1}{\sqrt{2}}\Bigl[\ket{\Psi_0}_\mathrm{NS}
+\ket{\Psi_0}_\mathrm{R}\Bigr],
\label{eq:initFV}
\end{equation}
where 
\begin{equation}
\ket{\Psi_0}_\mathrm{NS/R}=
\exp\left(\sum_{\substack{k\in\mathrm{NS/R}\\ k>0}}
K_\text{q}(\xi_k)\,A^\dagger(-\xi_k)\,A^\dagger(\xi_k)\right)
\ket{0}_\mathrm{NS/R}
\end{equation}
and $\ket{0}_\mathrm{NS/R}$ denotes the vacuum state in the corresponding sector.
In the regularization introduced by Fioretto and Mussardo (see Sec.~\ref{sec:FM}) the
quench matrix $K_\text{q}(\xi_k)$ has to be replaced by $K_{\tau_0}(\xi_k)$, which we will do 
in the numerical evaluations presented in Figs.~\ref{fig:fvreg1} and~\ref{fig:fvreg2}. 
The time evolution starting from the initial state \eqref{eq:initFV} now reads
\begin{equation}
\langle\sigma(t)\rangle=
\frac{2\;_\mathrm{NS}\!\bra{\Psi_0}\sigma(t)\ket{\Psi_0}_\mathrm{R}}
{_\mathrm{NS}\!\bra{\Psi_0}\Psi_0\rangle_\mathrm{NS}+
_\mathrm{R}\!\bra{\Psi_0}\Psi_0\rangle_\mathrm{R}}
=\sum_{m,n=0}^\infty D_{2m,2n}^L(t).
\label{eq:sigmaFV}
\end{equation}

Straightforward calculation of the norms gives
\begin{eqnarray}
_\mathrm{NS/R}\!\bra{\Psi_0}\Psi_0\rangle_\mathrm{NS/R}&=&
\sum_{n=0}^\infty Z_{\text{NS/R},2n}^L,\\
Z_{\text{NS/R},2}^L&=&\sum_{\substack{k,l\in\mathrm{NS/R}\\k,l>0}}K_\text{q}(\xi_k)^*\,
K_\text{q}(\xi_l)\;_\mathrm{NS/R}\!\bra{k,-k}-l,l\rangle_\text{NS/R}=
\sum_{\substack{k\in\mathrm{NS/R}\\k>0}}|K_\text{q}(\xi_k)|^2,\\
Z_{\text{NS/R},4}^L&=&\frac{1}{2}
\left(\sum_{\substack{k\in\mathrm{NS/R}\\k>0}}|K_\text{q}(\xi_k)|^2\right)^2
-\frac{1}{2}\sum_{\substack{k\in\mathrm{NS/R}\\k>0}}|K_\text{q}(\xi_k)|^4.
\end{eqnarray}
We note the analogy to the infinite-volume results \eqref{eq:appZ2} and \eqref{eq:appZ4}
respectively. The first non-trivial term in the expansion \eqref{eq:sigmaFV} reads
\begin{eqnarray}
D_{22}^L(t)&=&C_{22}^L(t)-\frac{\bar{\sigma}}{2}\left(Z_{\text{NS},2}^L+Z_{\text{R},2}^L\right)\\
C_{22}^L(t)&=&
\sum_{\substack{p\in\mathrm{NS}\\ p>0}}\;\sum_{\substack{q\in\mathrm{R}\\ q>0}}
K_\text{q}(\xi_p)^*\,K_\text{q}(\xi_q)\,
_\mathrm{NS}\!\bra{p,-p}\sigma\ket{-q,q}_\text{R}\,
e^{2M\ii t(\cosh\xi_p-\cosh\xi_q)}\\
&=&\frac{\bar{\sigma}}{(ML)^2}
\sum_{\substack{p\in\mathrm{NS}\\ p>0}}\;\sum_{\substack{q\in\mathrm{R}\\ q>0}}\;
K_\text{q}(\xi_p)^*\,K_\text{q}(\xi_q)\,\frac{\tanh\xi_p\,\tanh\xi_q}{\cosh\xi_p\,\cosh\xi_q}\,
\coth^2\frac{\xi_p-\xi_q}{2}\,\coth^2\frac{\xi_p+\xi_q}{2}\,e^{2M\ii t(\cosh\xi_p-\cosh\xi_q)}.
\qquad\label{eq:D22L}
\end{eqnarray}
In the same way a straightforward calculation yields
\begin{eqnarray}
D_{44}^L(t)&=&
\frac{\bar{\sigma}}{4(ML)^4}
\sum_{\substack{p_1,p_2\in\mathrm{NS}\\ p_1,p_2>0}}\;
\sum_{\substack{q_1,q_2\in\mathrm{R}\\ q_1,q_2>0}}\;
K_\text{q}(\xi_{p_1})^*\,K_\text{q}(\xi_{p_2})^*\,K_\text{q}(\xi_{q_1})\,K_\text{q}(\xi_{q_2})\,
\prod_{i=1}^2\frac{\tanh\xi_{p_i}\,\tanh\xi_{q_i}}{\cosh\xi_{p_i}\,\cosh\xi_{q_i}}\nonumber\\*
& &\hspace{20mm}\times
\tanh^2\frac{\xi_{p_1}+\xi_{p_2}}{2}\,\tanh^2\frac{\xi_{p_1}-\xi_{p_2}}{2}\,
\tanh^2\frac{\xi_{q_1}+\xi_{q_2}}{2}\,\tanh^2\frac{\xi_{q_1}-\xi_{q_2}}{2}
\nonumber\\*
& &\hspace{20mm}\times
\prod_{i,j=1}^2\coth^2\frac{\xi_{p_i}+\xi_{q_j}}{2}\,\coth^2\frac{\xi_{p_i}-\xi_{q_j}}{2}\,
\,e^{2M\ii t\sum_i(\cosh\xi_{p_i}-\cosh\xi_{q_i})}\nonumber\\*
& &-\frac{1}{2}\,C_{22}^L(t)\,\left(Z_{\text{NS},2}^L+Z_{\text{R},2}^L\right)
+\frac{\bar{\sigma}}{2}\,Z_{\text{NS},2}^L\,Z_{\text{R},2}^L
+\frac{\bar{\sigma}}{4}\,\left(Z_{\text{NS},4}^L+Z_{\text{R},4}^L\right).
\label{eq:D44L}
\end{eqnarray}
Finally, the relaxation rate \eqref{eq:relaxationrate} up to $\mathcal{O}(K_\text{q}^2)$ is
given by
\begin{equation}
\Gamma^L=
-\frac{2M}{L}\sum_{\substack{p\in\mathrm{NS}\\p>0}}|K_\text{q}(\xi_p)|^2\,\tanh\xi_p
-\frac{2M}{L}\sum_{\substack{q\in\mathrm{R}\\q>0}}|K_\text{q}(\xi_q)|^2\,\tanh\xi_q.
\end{equation}
We have evaluated \eqref{eq:D22L} and \eqref{eq:D44L} numerically using
the replacement $K_\text{q}\to K_{\tau_0}$ for several values of $\tau_0$, $L$ and $N$, 
where $N$ denotes the UV cut-off for the momentum numbers $p$ and $q$. 
The results are shown in Figs.~\ref{fig:fvreg1} and~\ref{fig:fvreg2}. We observe 
excellent agreement with the results obtained in the infinite volume. 

\end{document}